\pdfoutput=1
\documentclass[a4paper,11pt]{article}

\usepackage{graphicx}
\usepackage{amssymb}
\usepackage{amsmath}
\usepackage{fullpage}

\newcommand{\bra}[1]{\ensuremath{\langle #1|}}
\newcommand{\ket}[1]{\ensuremath{|#1\rangle}}
\newcommand{\ketbra}[2]{\ket{#1}\bra{#2}}
\newcommand{\fun}[2]{\ensuremath{#1\left(#2\right)}}
\DeclareMathOperator{\res}{\mathrm{Res}}
\DeclareMathOperator{\tr}{\mathrm{Tr}}
\newcommand{\Id}{{\cal I}}

\newcommand{\abs}[1]{\ensuremath{\left|#1\right|}}

\newcommand{\hermit}[1]{\ensuremath{{#1}^{\dagger}}}

\begin{document}

\title{General method for the security analysis in a quantum direct
communication protocol}
\author{Jaros{\l}aw Adam Miszczak\\Institute of Theoretical and Applied
Informatics, Polish Academy of Sciences,\\ Ba{\l}tycka 5, 44-100 Gliwice, Poland
\\Piotr Zawadzki\\Institute of Electronics,
Silesian University of Technology,\\Akademicka 16, 44-100 Gliwice, Poland}

\maketitle

\begin{abstract}
We introduce a general approach for the analysis of a quantum direct
communication protocol. The method is based on the investigation of the
superoperator acting on a joint system of the communicating parties and the
eavesdropper. The introduced method is more versatile than the approaches used
so far as it permits to incorporate different noise models in a unified way.
Moreover, it make use of a well grounded theory of quantum discrimination for
the purpose of estimating the eavesdropper's information gain.\\[6pt]
\emph{Keywords:} quantum direct communication, quantum cryptography
\end{abstract}

\section{Introduction}\label{sec:Introduction}
One of the main objectives of quantum information theory~\cite{NC00} is to
develop new protocols for controlling the networks of quantum processing units
connected by quantum channels~\cite{kimble08quantum}. This type of networks
requires new methods allowing full utilization of the capabilities offered by
quantum information processing~\cite{rass11unified}. Quantum internetworking
protocols should be able to exploit the quantum effects, including quantum
teleportation, dense coding and quantum direct communication. However, in order
to provide efficient methods for the utilization of quantum effects in large
scale networks, it is necessary to analyze the robustness of quantum information
processing against the errors that occur during the
transmission~\cite{lanzagorta12teleportation} and processing of quantum
information~\cite{gawron12noise}.

In this paper we focus on the analysis of the security of quantum ping-pong
protocol in the presence of a general form of quantum noise. The ping-pong
protocol~\cite{pz2372901:Bostrom:pingpong:2002} has attracted a lot of attention
as, contrary to quantum key distribution (QKD) schemes, it does not require
prior key agreement for confidentiality provision and it is provably
asymptotically secure in lossless channels. The theoretical success of the
protocol has been closely followed by the experimental implementation and the
proof of concept installation has been realized in the
laboratory~\cite{pz2372901:Ostermeyer:ping-pong-implementation:2008}. It has
also been shown that the protocol variant based on higher dimensional systems
and exploiting dense information coding also share features of the seminal
version when some improvements are
introduced~\cite{pz2372901:Vasiliu:qutrits:2011,pz2372901:zawadzki:qudits:2012}.

The ping-pong protocol, similarly to other quantum direct communication (QDC)
protocols, operates in two modes: a message mode is designed for information
transfer and a control mode is used for an eavesdropping detection. Although the
ping-pong protocol is asymptotically secure in perfect quantum channels, the
situation looks worse in noisy environments when legitimate users tolerate some
level of transmission errors and/or losses. If that level is too high compared
to the quality of the channel then an eavesdropper can peek some fraction of
signal particles hiding himself behind accepted quantum bit error rate (QBER)
threshold~\cite{pz2372901:wojcik:pingpong-attack:2003,pz2372901:zhang:pingpong:2004}.
Thus the possibility to intercept some part of the message without being
detected renders the protocol insecurity. To cope with this problem an
additional purely classical layer has been
proposed~\cite{pz2372901:zawadzki:ijqi:pingpong:2012}. However the estimation of
security improvement offered by that layer heavily depends on observed QBER and
the methods of the protocol analysis used so far do not offer mathematical
apparatus capable of QBER estimation in noisy channels. The purpose of the
presented work is to fill in this gap.

This paper is organized as follows.
In Section~\ref{sec:general-description} we provide a general description of the
ping-pong protocol in the language of density operators.
In Section~\ref{sec:noise-model} we describe a general model of noise which can
be used to describe the errors occurring during the execution of the protocol.
In Section~\ref{sec:passive} we apply the introduced description for the
purpose of the security in the passive eavesdropping scenario.
Finally, in Section~\ref{sec:concluding} we summarize the presented work and
provide concluding remarks.

\section{General description of the ping-pong protocol}\label{sec:general-description}
Let us consider the seminal version of the ping-pong
protocol~\cite{pz2372901:Bostrom:pingpong:2002} in which the message and control
modes are executed only in computational basis. The communication process is
started by Bob, the recipient of information, who prepares an EPR pair
\begin{equation}
\ket{\phi^+}=\left(\ket{0_B}\ket{0_A}+\ket{1_B}\ket{1_A}\right)/\sqrt{2}
\enspace.
\end{equation}
At the same time eavesdropping Eve controls her own system, which is initially
described by state~$\ket{\chi_E}$. As the states of Bob and Eve are separated,
the density matrix of the whole system reads
\begin{equation}
\label{pz2372901:eq:initial-density}
\rho_{BAE}^{(0)}=\rho_{BA}^{(0)}\otimes\rho_{E}^{(0)}=\ketbra{\phi^+}{\phi^+}\otimes\ketbra{\chi_E}{\chi_E}
\enspace .
\end{equation}
Next Bob sends a signal qubit $A$ to Alice. This qubit on its way can be
influenced by two factors: quantum noise because of channel imperfection and
malicious activities of Eve who may entangle it with the system controlled by
herself.

Let us assume that Eve is positioned close to Alice, so her action takes place
on the qubit modified by the noise. The density matrix of the system just before
signal qubit enters the environment controlled by Alice reads
\begin{equation}
\label{pz2372901:eq:after-noise}
\rho_{BAE}^{(1)}
= \fun{\left({\cal N}_{BA}\otimes \Id_E\right)}{\rho_{BA}^{(0)}\otimes\rho_{E}^{(0)}}
= \rho_{BA}^{(1)}\otimes \rho_{E}^{(0)}
\enspace ,
\end{equation}
where it has been explicitly highlighted that noise operator $\cal N$ acts only
on the EPR pair ($\cal I$ denotes identity operation). Before signal qubit
enters Alice's environment, Eve can entangle it with her own system
\begin{equation}
\label{pz2372901:eq:after-entangle}
\rho_{BAE}^{(2)}
= \left(\Id_B \otimes {\cal E}_{AE}\right)\rho_{BAE}^{(1)}
\enspace ,
\end{equation}
where entangling operator ${\cal E}_{AE}$ acts only on qubit~$A$ of the EPR pair
and system possessed by Eve. At that point of protocol execution Alice can
select a control mode which serves for eavesdropping detection or continue in
the information mode.

In the former case she measures the received qubit in computational basis,
\emph{i.e.} performs von Neumann measurement using projectors
$M_{x,A}=\Id_B\otimes \ketbra{x_A}{x_A}\otimes \Id_E$, $x=0,1$. The probability
that she finds qubit under investigation in state $\ket{x}$ (measures $\pm 1$)
is given by
\begin{equation}
\fun{p_A}{x} = 
\fun{\tr}{
\rho_{BAE}^{(2)} 
M_{x,A}
}
\end{equation}
After the measurement the state of the whole system is described by 
\begin{equation}
{\sigma_x}_{BAE}^{(2)}
=
\frac{M_{x,A} \rho_{BAE}^{(2)} M_{x,A}}{ \fun{\tr}{\rho_{BAE}^{(2)}M_{x,A}} }. 
\end{equation}
Subsequently Bob measures his qubit in computational basis using projectors
$M_{y,B}=\ketbra{y_B}{y_B}\otimes \Id_A\otimes \Id_E$, $y=0,1$. The probability
that Bob finds his qubit in state $\ket{y}$ provided that Alice has found his
qubit in state $\ket{x}$ is given by
\begin{equation}
\fun{p_{B|A}}{y|x} = \fun{\tr}{{\sigma_x}_{BAE}^{(2)}  M_{y,B}}.
\end{equation} 
From the above, it follows that errors in control mode appear with probability
\begin{equation}
\label{pz2372901:eq:control-mode-failure-def}
P_{EC} = \fun{p_{B|A}}{1|0} \fun{p_A}{0} + \fun{p_{B|A}}{0|1} \fun{p_A}{1}.
\end{equation}

In information mode, Alice encodes a classic bit $\mu$ applying ($\mu=1$) or not
($\mu=0$) operator $Z_A$ to the possessed qubit. The system state after encoding
is given by
\begin{equation}
\label{pz2372901:eq:states-after-encoding}
{\rho_\mu}_{BAE}^{(3)}
=
\left(\Id_B\otimes Z_A^\mu\otimes \Id_E\right)
\rho_{BAE}^{(2)}
\left(\Id_B\otimes \left(Z_A^\mu\right)^\dag\otimes \Id_E\right)
\enspace .
\end{equation}
The qubit $A$ is sent back to Bob after the encoding operation.
Eve's task is to discriminate
between states 
${\rho_\mu}_{AE}^{(3)}=\fun{\tr_B}{{\rho_\mu}_{BAE}^{(3)}}$
with
maximal confidence.
The system states after Bob's reception of qubit~$A$ traveling back from Alice
and in the absence of Eve measurements are given by
\begin{equation}
\label{pz2372901:eq:after-enconding-and-noise}
{\rho_\mu}_{BAE}^{(4)}
= 
\left({\cal N}_{BA}\otimes \Id_E\right)
{\rho_\mu}_{BAE}^{(3)},
\end{equation}
so Bob has to distinguish the states
\begin{equation}
\label{pz2372901:eq:Bob-after-enconding-and-noise}
{\rho_\mu}_{BA}^{(4)}=\fun{\tr_E}{{\rho_\mu}_{BAE}^{(4)}}
\enspace .
\end{equation}

When Eve performs measurements,
the same quantum discrimination strategy is used
but Bob is unconscious that measured states are of the form
\begin{equation}
{\tau_{\mu,\alpha}}_{BA}^{(4)}
=
\fun{\tr_E}{
\left({\cal N}_{BA}\otimes \Id_E\right)
\frac{
M_{\alpha,E} {\rho_\mu}_{BAE}^{(3)} M_{\alpha,E}
}{
\fun{\tr}{{\rho_\mu}_{BAE}^{(3)} M_{\alpha,E}} 
}
}.
\end{equation} 

The analysis of the protocol should determine Eve's information gain $I_E$ and
the probability of erroneous Bob's decoding $QBER$ as functions of probability
of error observed in control mode $P_{EC}$ and, optionally, parameters
describing noise operator ${\cal N}$.

\section{Model of the noise}\label{sec:noise-model}
Any interaction with the environment observed from the perspective of the
principal system can be given as operator sum (Kraus)
representation~\cite{NC00,BP07}
\begin{equation}
\label{pz2372901:eq:quantum-channel}
\rho\to \rho^\prime = {\cal N}\rho = \sum_k K_k \rho \hermit{K}_k
\end{equation}
provided that $\sum_k K_k \hermit{K}_k = \Id$.
Such an approach hides the details of the interaction of the system under
investigation with the environment, but these details are not of immediate
relevance in analysis of many quantum information processing related tasks. In
such situations Kraus representation proved to be useful because it provides a
unified description of many, seemingly different, physical processes.

It follows from the description of the protocol operation that the signal qubit
only interacts with the environment. If that qubit was separated from the home
qubit possessed by Bob, the noise operator would be simply decomposed as
${\cal N}_{BA}= {\Id}_B \otimes {\cal N}_A $, where ${\cal N}_A$ describes the
interaction of the signal qubit with the environment. 
However, the signal and
home qubits are maximally entangled in the ping-pong protocol and such
decomposition is not obvious. In this section we introduce a general procedure
for deriving Kraus operator for such a situation and exemplify it with the
depolarizing channel.

The action of the quantum channel~\eqref{pz2372901:eq:quantum-channel} can be
represented as a supermatrix~$M_{\cal N}$~\cite[Eq.~(5)]{pz2372901:Miszczak:singular:2011}
\begin{equation}
\fun{\res}{\rho^\prime}= \sum_k \fun{\res}{K_k \rho \hermit{K_k}}
= \sum_k K_k^\star\otimes K_k \fun{\res}{\rho}
= M_{\cal N} \fun{\res}{\rho},
\end{equation}
where $\fun{\res}{\cdot}$ denotes reshape transformation which maps a density
matrix into a~column vector row wise. The Kraus operators can be recovered from
$M_{\cal N}$ with singular matrix decomposition of the matrix $D_{\cal N}$ which
is related to $M_{\cal N}$ via
\begin{equation}\label{pz2372901:eq:dynamical-matrix}
M_{\cal N} = \sum_{k,l} \left\{D_{\cal N}\right\}_{k,l} \epsilon_k \otimes \epsilon_l,
\end{equation}
where $\epsilon_k$ denotes $k$-th element of the canonical base~of $\rho$
space~\cite{pz2372901:Miszczak:singular:2011}. When two quantum channels ${\cal
N}_B$ and ${\cal N}_A$ are applied to the parts of the composite system the
supermatrix of the composite channel can be found as
\begin{equation}\label{pz2372901:eq:prod-super-mtx}
M_{{\cal N}_B\otimes {\cal N}_A} = M_R({\cal N}_B\otimes {\cal N}_A)M_R,
\end{equation}
where $M_R$ is the matrix representing the change of base in the product space
and it is commonly called the reshuffle
matrix~\cite{pz2372901:Miszczak:singular:2011,pz2372901:bengtsson:2006:geometry}.
In the considered case ${\cal N}_B \equiv \Id_B$ and our task is to derive Kraus
operators for $M_{\Id_B\otimes{\cal N}_A}$ and subsequently a map for density
operators analogous to~\eqref{pz2372901:eq:quantum-channel}.

The single qubit depolarizing channel with reliability $r$, which is commonly
used to model white noise, is described by the map~\cite{NC00} 
\begin{equation}
{\cal N}_A^{(D)}(\rho_A) 
              = r \rho_A + \frac{1-r}{4} \sum_{k=0}^3  \sigma_k \rho_A \hermit{\sigma_k},
\end{equation}
where $\sigma_k$ are Pauli matrices.
The supermatrix of this map is given as
\begin{equation}
M_{{\cal N}_A}^{(D)} = \left( \begin{smallmatrix}
 \frac{1+r}{2} & 0 & 0 & \frac{1-r}{2} \\
 0 & r & 0 & 0 \\
 0 & 0 & r & 0 \\
 \frac{1-r}{2} & 0 & 0 & \frac{1+r}{2} \\
\end{smallmatrix}\right).
\end{equation}
Using the formula \eqref{pz2372901:eq:prod-super-mtx}, one gets the explicit form of the
supermatrix for the extended channel
\begin{equation}\label{pz2372901:eq:supermatrix-depolar}
M^{(D)}_{\Id_B\otimes {\cal N}_A} = \left[
\begin{smallmatrix}
 \frac{1+r}{2} & 0 & 0 & 0 & 0 & \frac{1-r}{2} & 0 & 0 & 0 & 0 & 0 & 0 & 0 & 0 & 0 & 0 \\
 0 & r & 0 & 0 & 0 & 0 & 0 & 0 & 0 & 0 & 0 & 0 & 0 & 0 & 0 & 0 \\
 0 & 0 & \frac{1+r}{2} & 0 & 0 & 0 & 0 & \frac{1-r}{2} & 0 & 0 & 0 & 0 & 0 & 0 & 0 & 0 \\
 0 & 0 & 0 & r & 0 & 0 & 0 & 0 & 0 & 0 & 0 & 0 & 0 & 0 & 0 & 0 \\
 0 & 0 & 0 & 0 & r & 0 & 0 & 0 & 0 & 0 & 0 & 0 & 0 & 0 & 0 & 0 \\
 \frac{1-r}{2} & 0 & 0 & 0 & 0 & \frac{1+r}{2} & 0 & 0 & 0 & 0 & 0 & 0 & 0 & 0 & 0 & 0 \\
 0 & 0 & 0 & 0 & 0 & 0 & r & 0 & 0 & 0 & 0 & 0 & 0 & 0 & 0 & 0 \\
 0 & 0 & \frac{1-r}{2} & 0 & 0 & 0 & 0 & \frac{1+r}{2} & 0 & 0 & 0 & 0 & 0 & 0 & 0 & 0 \\
 0 & 0 & 0 & 0 & 0 & 0 & 0 & 0 & \frac{1+r}{2} & 0 & 0 & 0 & 0 & \frac{1-r}{2} & 0 & 0 \\
 0 & 0 & 0 & 0 & 0 & 0 & 0 & 0 & 0 & r & 0 & 0 & 0 & 0 & 0 & 0 \\
 0 & 0 & 0 & 0 & 0 & 0 & 0 & 0 & 0 & 0 & \frac{1+r}{2} & 0 & 0 & 0 & 0 & \frac{1-r}{2} \\
 0 & 0 & 0 & 0 & 0 & 0 & 0 & 0 & 0 & 0 & 0 & r & 0 & 0 & 0 & 0 \\
 0 & 0 & 0 & 0 & 0 & 0 & 0 & 0 & 0 & 0 & 0 & 0 & r & 0 & 0 & 0 \\
 0 & 0 & 0 & 0 & 0 & 0 & 0 & 0 & \frac{1-r}{2} & 0 & 0 & 0 & 0 & \frac{1+r}{2} & 0 & 0 \\
 0 & 0 & 0 & 0 & 0 & 0 & 0 & 0 & 0 & 0 & 0 & 0 & 0 & 0 & r & 0 \\
 0 & 0 & 0 & 0 & 0 & 0 & 0 & 0 & 0 & 0 & \frac{1-r}{2} & 0 & 0 & 0 & 0 & \frac{1+r}{2}\\
\end{smallmatrix}
\right].
\end{equation}
Kraus operators for the above extended channel can be obtained using
eigendecomposition of the dynamical matrix~\eqref{pz2372901:eq:dynamical-matrix} corresponding to the
supermatrix~$M^{(D)}_{ \Id_B \otimes{\cal N}_A}$:
\begin{multline}\label{pz2372901:eq:kraus-canonical}
K^{(D)}_{ \Id_B \otimes {\cal N}_A}=\left\{
-\frac{\sqrt{1-r}}{2}\,
\Id_B \otimes \sigma_z
,
\frac{\sqrt{1-r}}{\sqrt{2}}\, \Id_B \otimes
\left[\begin{smallmatrix} 0 & 1 \\ 0 & 0 \end{smallmatrix}\right]
,
\right. \\ \left.
\frac{\sqrt{1-r}}{\sqrt{2}}\, \Id_B \otimes
\left[\begin{smallmatrix} 0 & 0 \\ 1 & 0 \end{smallmatrix}\right]
,
\frac{\sqrt{1+3 r}}{2}\, \Id_B \otimes \Id_A
\right\}.
\end{multline}
Kraus representation \eqref{pz2372901:eq:kraus-canonical} implies that the
action of the extended channel on the second subsystem is independent from the
action on the first subsystem. Using the representation of the composite system
density matrix
\begin{multline}
\rho_{BA} 
=
\left[
\begin{smallmatrix}
\rho_{0_B 0_A,0_B 0_A} & \rho_{0_B 0_A,0_B 1_A} & \rho_{0_B 0_A,1_B 0_A} & \rho_{0_B 0_A,1_B 1_A} \\
\rho_{0_B 1_A,0_B 0_A} & \rho_{0_B 1_A,0_B 1_A} & \rho_{0_B 1_A,1_B 0_A} & \rho_{0_B 1_A,1_B 1_A} \\
\rho_{1_B 0_A,0_B 0_A} & \rho_{1_B 0_A,0_B 1_A} & \rho_{1_B 0_A,1_B 0_A} & \rho_{1_B 0_A,1_B 1_A} \\
\rho_{1_B 1_A,0_B 0_A} & \rho_{1_B 1_A,0_B 1_A} & \rho_{1_B 1_A,1_B 0_A} & \rho_{1_B 1_A,1_B 1_A} \\
\end{smallmatrix}
\right]
= \\
=
\sum\limits_{m,n=\{0,1\}}
\left[\tr_A \rho_{BA}\right]_{mn}
\ketbra{m_B}{n_B} \otimes
\left(
\frac{1}{\left[\tr_A \rho_{BA}\right]_{mn}}
\left[
\begin{matrix}
\rho_{m_B 0_A,n_B 0_A} & \rho_{m_B 0_A,n_B 1_A} \\
\rho_{m_B 1_A,n_B 0_A} & \rho_{m_B 1_A,n_B 1_A} \\
\end{matrix}
\right]
\right)
\end{multline}
one can show that it implies the following map
\begin{equation}\label{pz2372901:eq:composite-map}
{\cal N}_{BA} \rho_{BA} =
\left(\Id_B \otimes {\cal N}_{A} \right) \rho_{BA} =
r \rho_{BA} + \frac{1-r}{2} \tr_{A}(\rho_{BA})\otimes \Id_A .
\end{equation}

\section{Passive eavesdropping in a noisy channel}\label{sec:passive}
As the application of the introduced description, we consider a simple model of
quantum noise described by the depolarizing channel.
Let us consider the situation in which Eve does not entangle with a signal qubit 
i.e. ${\cal E}_{AE}=\Id_{AE}$.
Such assumption results in the separation of the system controlled by Eve, 
so it is not taken into account in further expressions.
From the map~\eqref{pz2372901:eq:composite-map} it follows that the system state
after the reception of the signal qubit by Alice reads 
\begin{multline}
\rho_{BA}^{(2)}=\frac{r}{2}
\left[
\ketbra{0_B0_A}{1_B1_A}+
\ketbra{1_B1_A}{0_B0_A}
\right] + \\
+ \frac{1-r}{4}
\left[
\ketbra{1_B0_A}{1_B0_A}+
\ketbra{0_B1_A}{0_B1_A}
\right] + \\
+ \frac{1+r}{4}
\left[
\ketbra{0_B0_A}{0_B0_A}
+\ketbra{1_B1_A}{1_B1_A}
\right].
\end{multline}
It follows from \eqref{pz2372901:eq:control-mode-failure-def} that the
probability of error occurrence in control mode is equal to
\begin{equation}
\label{pz2372901:eq:control-mode-failure}
P_{EC}=(1-r)/2.
\end{equation}

In the information mode the encoding operation leads to states
\begin{multline}
{\rho_\mu}_{BA}^{(3)}=
(-1)^\mu
\frac{r}{2}
\left[
\ketbra{0_B0_A}{1_B1_A}+
\ketbra{1_B1_A}{0_B0_A}
\right] + \\
+ \frac{1-r}{4}
\left[
\ketbra{1_B0_A}{1_B0_A}+
\ketbra{0_B1_A}{0_B1_A}
\right] + \\
+ \frac{1+r}{4}
\left[
\ketbra{0_B0_A}{0_B0_A}
+\ketbra{1_B1_A}{1_B1_A}
\right].
\end{multline}
However, Eve's observation capabilities are limited only to travelling qubit
$A$, thus she has to distinguish between states
${\rho_\mu}_{A}^{(3)}=\fun{\tr_B}{{\rho_\mu}_{BA}^{(3)}}$. But
${\rho_0}_{A}^{(3)}={\rho_1}_{A}^{(3)}$ and, in consequence, $I_E=0$. Thus
the~noise gives no additional advantage to passively eavesdropping Eve.

The qubit $A$ in its way back to Bob again interacts with the environment so he
receives the states
\begin{multline}
{\rho_\mu}_{BA}^{(4)}  = r {\rho_\mu}_{BA}^{(3)} + \frac{1-r}{2} \fun{\tr_A}{{\rho_\mu}_{BA}^{(3)}} \otimes \Id_A 
= \\ =
(-1)^\mu
\frac{r^2}{2}
\left[
\ketbra{0_B0_A}{1_B1_A}+
\ketbra{1_B1_A}{0_B0_A}
\right] + \\
+ \frac{1-r^2}{4}
\left[
\ketbra{1_B0_A}{1_B0_A}+
\ketbra{0_B1_A}{0_B1_A}
\right] 
+ \\ + \frac{1+r^2}{4}
\left[
\ketbra{0_B0_A}{0_B0_A}
+\ketbra{1_B1_A}{1_B1_A}
\right].
\end{multline}
\begin{figure}
\centering
    \includegraphics[width=\textwidth]{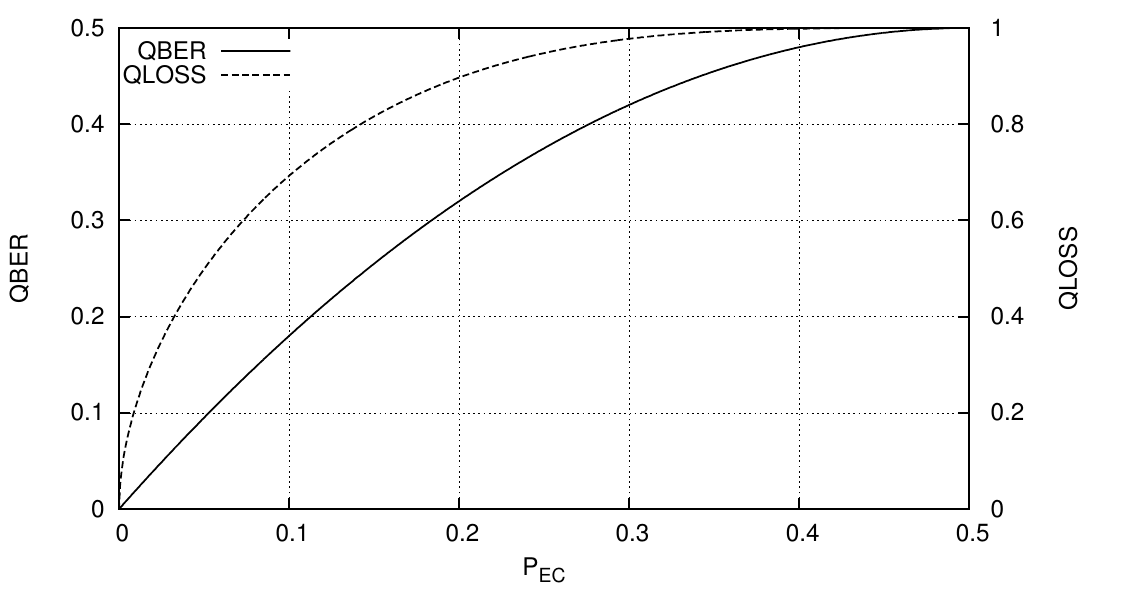}
    \caption{Probabilities of a~particle loss ($QLOSS$) or an~erroneous decoding
    ($QBER$) as a function of control mode failure probability ($P_{EC}$) in
    protocol operation over depolarizing channel.}
\label{pz2372901:fig:qber-vs-qloss}
\end{figure}

If Bob uses unambiguous discrimination, the bits are lost (measurement fails) 
with a~probability~\cite{pz2372901:Herzog:unambiguous:2004}
\begin{equation}
\label{pz2372901:eq:Bob-QLOSS}
QLOSS = 1 - P_s^\mathrm{max} = 
\fun{F}{{\rho_0}_{BA}^{(4)},{\rho_1}_{BA}^{(4)}},
\end{equation}
where the overlap of ${\rho_\mu}_{BA}^{(4)}$ is given by
\begin{equation}
\fun{F}{{\rho_0}_{BA}^{(4)},{\rho_1}_{BA}^{(4)}}=
\frac{1}{2}(1 - r^2) + \frac{1}{2} \sqrt{\left(1 - r^2\right)\left(1 + 3r^2\right)}.
\end{equation}
On the other hand, if Bob uses minimum error discrimination the observed
bit error rate is equal to~\cite{pz2372901:Fuchs:distinguisability:99}
\begin{equation}
\label{pz2372901:eq:Bob-QBER}
QBER
= \frac{1}{2}\left(
1-\frac{1}{2}\fun{\tr}{\abs{{\rho_0}_{BA}^{(4)}-{\rho_1}_{BA}^{(4)}}}
\right)
= (1-r^2)/2.
\end{equation}
Quantities $QBER$ and $QLOSS$ as a function of control mode failure
probability~\eqref{pz2372901:eq:control-mode-failure-def}, which is a parameter
directly accessible to communicating parties, are shown in
\figurename~\ref{pz2372901:fig:qber-vs-qloss}. Both $QBER$ and $QLOSS$ do not
scale linearly with $P_{EC}$. Moreover, the functional form of the obtained
scaling heavily depends on the parameters of the noise model used, thus the
correct modeling of noise is of prime importance in the estimation of the
protocol operation over non-perfect quantum channels.

\section{Concluding remarks}\label{sec:concluding}

The usefulness of the general method based on density operator analysis for
ping-pong protocol operation has been presented. As the proof of concept the
example of its application to the analysis of the protocol execution over
depolarizing channel has been given. The analysis of a~more complicated case of
an~active eavesdropping is left for future research. Although the~method is more
cumbersome than the approach used so far, it is more versatile as it permits an
incorporation of different models of noise in a unified way and makes use of
a~well grounded theory of quantum discrimination in estimation of eavesdropper's
information gain.

\section*{Acknowledgements}
Authors would like to acknowledge the support by the Polish National Science
Centre under the research project UMO-2011/03/D/ST6/00413.

\end{document}